\documentclass[twocolumn]{rich2004}

 \usepackage{epsfig}

\usepackage{amssymb}

\begin{document}

\begin{frontmatter}



\title{The BTeV RICH Front End Electronics}


\author[syr]{Marina Artuso}
\centerline{{\it on behalf of the BTeV RICH Group}}

\address[syr]{Department of Physics, Syracuse University, Syracuse, NY 12344, USA}

\begin{abstract}
We report on the design and testing of novel mixed analog and
digital front end ASICs custom made for the single photon
detectors considered for the BTeV RICH system. The key features
are reviewed, as well as results achieved using electronics bench
tests and  beam studies.
\end{abstract}

\begin{keyword}
BTeV \sep Front End Electronics \sep Photodetectors
\PACS  85.40.-e \sep 85.60 Gz \sep 85.60 Bt \sep 07.07df
\end{keyword}
\end{frontmatter}

\section{Introduction}
Charged particle identification is a crucial component of any
modern experiment studying charm and beauty decays. Ring Imaging
Cherenkov (RICH) detectors offer a very attractive approach. The
BTeV RICH detector has the important functions of separating
$\pi,\ K,\ {\rm and}\ p$, and also discriminating electrons and
muons from pions.

The BTeV RICH detector is designed to separate $\pi/K/p$ in a
momentum range of 3 to 70 GeV/$c$~\cite{btev:tdr}. It is essential
to CP violation studies, providing separation of specific final
states, such as $K^+\pi^-$ from $\pi^+\pi^-$, and also providing
information on the $b$ flavor by identifying charged kaons. The
RICH detector also complements the electromagnetic calorimeter and
the muon detector in lepton identification and thus it increases
the reconstruction efficiency in interesting modes like $B^0
\rightarrow J/\psi K_s$ significantly. We use a novel gas radiator
($\rm C_4 F_8 O)$ \footnote{$\rm C_4 F_8 O$ is produced by 3M,
USA, http://www.3m.com/} to generate Cherenkov light in the
optical frequency range. The light is focused by mirrors onto two
photon detector arrays. The two photon detector options that we
considered were hybrid photo-diodes (HPDs), fabricated at DEP, the
Netherlands \footnote{Delft Electronic Products B.V. (DEP),
http://www.dep.nl/} ,  and multi-anode photomultiplier tubes
(MaPMTs), fabricated by Hamamatsu\footnote{Hamamatsu Photonics,
Japan; http://usa.hamamatsu.com/}, Japan. To separate kaons from
protons below the threshold of gaseous radiator, a liquid radiator
($\rm C_ 5 F_{12}$) is used and the light, proximity focused onto
the sides of the vessel, is detected by 3" PMT arrays.

The design of these devices poses interesting
challenges. Different tradeoffs between dynamic range and low
intrinsic noise play a role in achieving optimal performance.
Timing requirements are equally important, as we need to store the
time-stamped event information in local buffers prior to the next
collision, assumed to occur within 132 ns. Finally, although the
RICH detector has generally very low occupancy, some areas
experience high rates and we required these devices to be able to
sustain the maximum rate, expected to be on the order of 3 MHz.

Two different custom made ASICs were designed and  produced for us
by IDEAS, Norway \footnote{Ideas ASA, Ideas ASA, N-1330 Fornebu,
Norway; http://www.ideas.no.} The most extensive tests were
performed on the VA\_MaPMT circuit, packaged in the hybrid devices
used in the BTeV RICH beam test studies described in another
contribution to this conference \cite{tomasz}.

\section{Conceptual design of the VA\_MaPMT ASICs}
The detector segmentation chosen for our system makes it natural
to use a binary output, as the signal is expected to be contained
within a single pixel cell and the occupancy is very low so that
it is very unlikely that a pixel is hit by more than 1 photon
within any given event. This choice minimizes the number of bits
of the hit information to be transferred. Finally, as the data
needs to be collected in real time, the front end electronics
includes self-triggering capabilities.

The specifications of the analog processor are different for the
two photon detectors. The hybrid photon detectors (HPDs) pose the
most stringent requirements on the intrinsic and common mode noise
of the readout system. The expected charge signal distribution has
a narrow width centered at about 5,000 $e^-$; thus the equivalent
noise charge (ENC) of the electronics processor should not exceed
500 $e^-$. Moreover, as the front end electronics incorporates a
built-in discriminator, common mode noise must be negligible. This
is a significant challenge for a device with distributed readout
electronics and a 20 KV high voltage system. On the other hand,
the MaPMT tubes feature an average gain in excess of 10$^6$. Even
though the single photon response in this case is rather broad,
the noise requirements can be relaxed, but high dynamic range
becomes a critical feature.

\begin{figure*}[hbt]
\center{\vspace{-1in} \epsfig{figure = 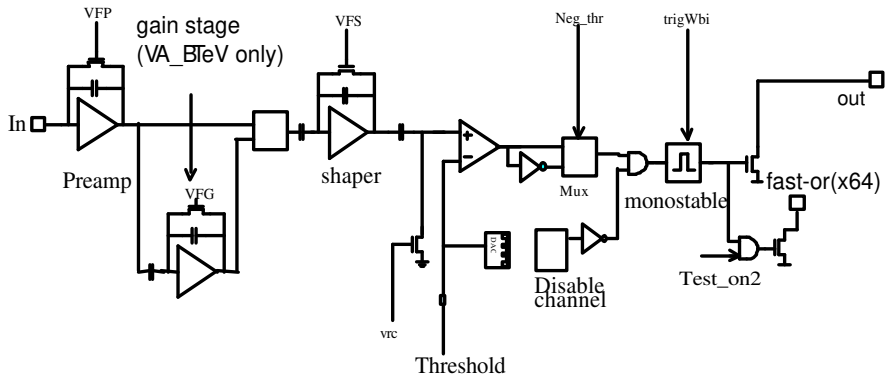,width=5in}
}\vspace{-0.5in} \caption{ Block diagram of an individual readout
channel of the front end ASICs described in this paper.}
\label{fig:bldia}
\end{figure*}

These devices feature 64 parallel inputs and 64 parallel outputs.
The parallel output architecture allows a prompt extraction of the
channel hit information and the peripheral electronics can attach
a time stamp to each event. The slow control is implemented as a
serial bitstream that programs the mode of operation of the ASIC
and allows the fine tuning of the individual channel thresholds
described below.

Fig. 1 shows the conceptual diagram of each readout channel. The
analog section comprises a semi-Gaussian preamplifier and shaper
circuit, followed by a high pass filter that reduces the
discriminator sensitivity  to long range drifts of the DC working
point of the device. In addition, a voltage-controlled pole-zero
cancellation circuit is introduced to optimize the rate
capabilities.

The input of the digital section is a discriminator that must
operate effectively at very low thresholds and it needs to
tolerate very high rates, of the order of several MHz, to cope
with the high occupancy expected in some areas. The discriminator
threshold is set through an external 8 bit DAC. In addition, a 4
bit programmable DAC is built in every channel to fine tune the
threshold of each individual channel to compensate for different
DC offsets. The discriminator output drives a monostable circuit
that produces an output current pulse whose width is about 100 ns.
Individual digital outputs can be disabled through a channel mask
set during the initialization sequence.

There are three modes of operation for this ASIC: (1) an
initialization sequence, when a bit pattern sequence is shifted in
the ASIC to program the desired operating conditions; (2) a
calibration mode, when channels selected in the initialization
sequence respond to an input current pulse sent to the calibration
input; (3) finally, in normal mode, all the working channels are
activated and respond to charge signals collected at their inputs.
In addition, a fast-OR of all the channel hits can be activated
for monitoring or synchronization purposes.

\section{The front end ASICs}
We started our R\& D work with the ASICs best matched to the HPD application
(VA\_BTeV). In order to operate the
discriminator with threshold levels of the order of about 30 mV, the
RC-CR shaper is complemented by an optional gain stage, which
provides an additional 3-fold amplification. When proper shielding
and grounding for this device was achieved, the expected ENC of
500 $e^-$ at about 10 pF input capacitance was achieved and we
were able to see efficient response to our blue LED single
photon source.

Our next project involved the development of an ASIC to be used in
conjunction with the R8900-M16 MaPMT tubes. In this version
(VA\_MaPMT), we reduced the gain of the analog front end, but we
focused the design primarily in maintaining the capability of
operating at very low thresholds. The linear range of the
analog front end extended to 220 fC, assumed to be adequate for this
application. An interesting
feature added to this design was an analog channel that can be
used to monitor the analog front end response either to a
calibration pulse or can be connected to a given MaPMT pixel. This
control channel proved extremely valuable in understanding the
performance of this system with different ASIC biasing conditions
and MaPMT high voltages. The performance of these ASICs will be
the focus of this paper, as they have been more extensively tested
in a variety of conditions.

The devices described in this paper are implemented in the 0.8
$\mu$m N-well CMOS AMS process. A new iteration has been
implemented in the 0.35 $\mu$m CMOS process and features an
extended dynamic range for MaPMT applications.

\section{The front end hybrids}

\begin{figure}
\center{  \epsfig{figure =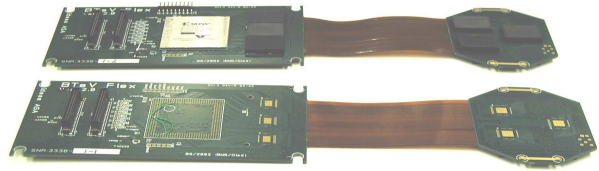,height=1.in} }
\caption{VA\_BTeV hybrid implemented on rigid-flex technology.}
\label{flex-rigid}
\end{figure}

The HPD application involved also a packaging challenge. The 163
pixel HPDs produced for us by DEP \cite{dep} had the pixel output
brought outside the HPD vacuum by a pin array. This arrangement
did not leave much space for electronics components, thus we used
a rigid-flex-rigid technology, shown in Fig.~\ref{flex-rigid}. The
octagonal rigid section contains the VA\_BTeV ASICs and their
biasing circuits, while the digital back end, coordinating the
data transfer from front end ASICs to the data
acquisition system, is hosted on the bigger
rectangular section. The two rigid elements were connected
electrically by a flex circuit embedded in the two rigid planes,
that allowed the positioning of the digital section at a right
angle with respect to the photon array plane.

\begin{figure}
\center{ \vspace{-1.5in}\epsfig{figure =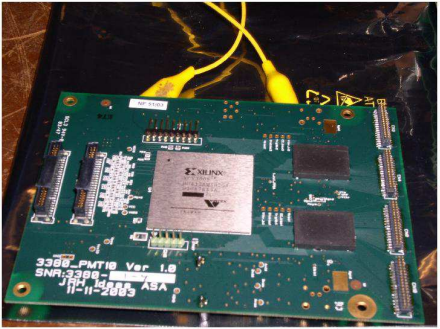,width=2.5
in} } \caption{ Va\_MaPMT ASICs mounted on the hybrids used in the
BTEV gas RICH prototype studied with test beam runs at Fermilab.}
\label{mapmt-hyb}
\end{figure}

Fig.~\ref{mapmt-hyb} shows the hybrid hosting the VA\_MaPMT ASICs.
It is a conventional 6-layer rigid printed circuit board. The
analog inputs are routed via flat multiconductor cables to the
base board hosting the MaPMT photon detectors and providing the
high voltage biasing network for their 12 stages.

Both hybrids incorporate level shifter circuits that translate the
current output from the front end ASICs into voltage level
matching the input requirements of the XILINX Virtex 300 FPGA,
used to drive the initialization sequence and to latch and
transfer the data from the front end to the back end circuit with
the protocols needed by the data acquisition system. The
firmware can be downloaded in the first step of the initialization
sequence and thus we could adapt this hybrid to different data
taking modes, and different triggering configurations.

\section{Performance}
The initial set of measurements
involved input charge scans at different threshold voltages.
The shape of the transition between zero and maximal efficiency
was used to estimate the intrinsic noise of the ASIC under test.
From these measurements we inferred that a typical noise for the
VA\_BTeV mounted on the hybrid described above is about 500 $e^-$,
whereas the typical noise for the VA\_MaPMT mounted on their
custom made hybrids is about 2,000 $e^-$.

The next tests involved the hybrids attached to their
corresponding photon detector. The assembly was located in a test
box where light from a blue LED was collimated onto a pixel of the
photon detector. The LED was driven by a narrow pulse
generator that biased the LED at tuneable very low light level.
Both detector-electronics pairs demonstrated very good single
photon sensitivity.

A set of ten MaPMT hybrids were used in a gas radiator RICH
prototype that was taken to the test beam facility at Fermilab.
Data were taken in two different periods, separated by a few
months.

The results of the first set of data are discussed in a separate
contribution to this conference \cite{tomasz}.  We were able to
operate at very low threshold throughout the duration of the data
taking without additional noise hits. In particular, we were able
to run with thresholds of 5 mV, corresponding to a minimum charge
signal of 27,000 e$^-$. With this threshold, the ratio between the
input charge producing the onset of saturation and the minimum
input signal to trigger the discriminator response is 52.

\begin{figure*}
\center{ \epsfig{figure =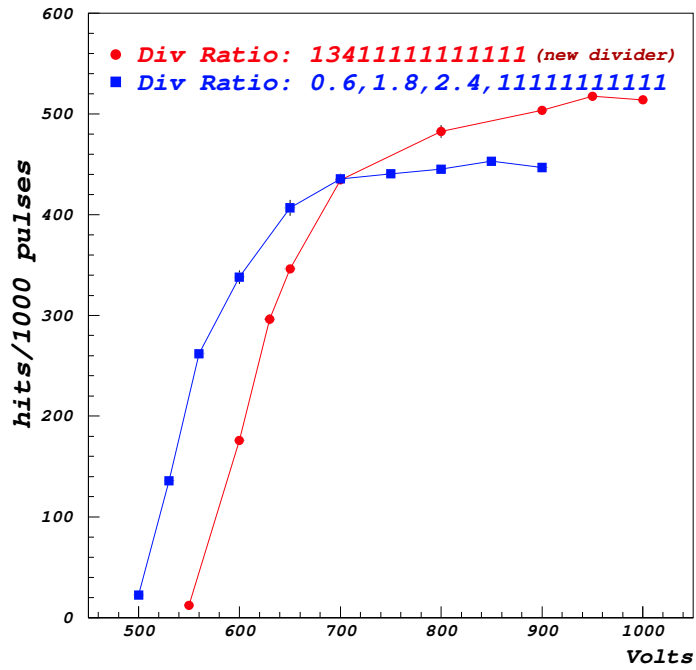,width=4in} }
\center{\epsfig{figure =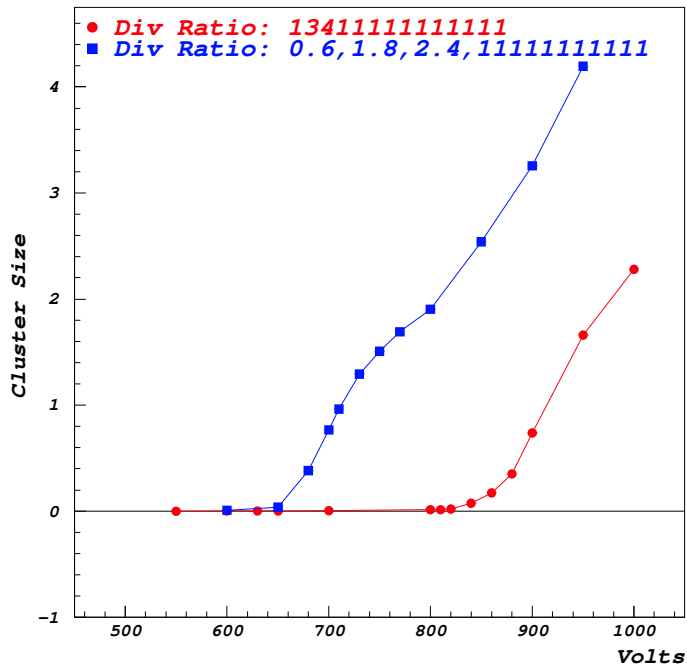,width=4 in} }

\caption{Comparison of the performance of the front end
electronics with the old and new voltage divider ratio for the
R8900 MaPMT bias: (top) efficiency (arbitrary units) versus high
voltage, (bottom) cross talk measured in the two configurations,
the vertical axis is defined in the text. The voltage dividers are
identified by the ratio between the resistors in the biasing
chain.} \label{crosstalk}
\end{figure*}

During the first run we observed unexpected cross talk between
neighboring channels when the high voltages exceeded the onset of
the plateau by about 50 to 100 V. Studies on the test channel with
analog output connected to a  MaPMT tube showed that the major
cause of the cross talk was related to the analog channel going
into saturation more and more frequently as the high voltage was
increased above plateau. In order to reduce the gain without any
collection efficiency loss at the first dynode, a different
biasing condition for the MaPMT was suggested by Hamamatsu
\cite{hama:private}. This new voltage divider ratio allows for the
minimum voltage between the anode and the first dynode that is
needed to achieve full collection efficiency at a lower value of
the tube gain. The tradeoff between gain and crosstalk in the two
configurations is shown in Fig.~\ref{crosstalk}. The cross talk is
characterized by the average number of hits recorded among the 6
closest neighbors to the channel attached to the MaPMT pixel: a
value of 0 means that there is no cross talk, a value of 6 means
that the neighboring channels are registering a hit whenever the
MaPMT pixel is hit. The new biasing scheme represents an
improvement, but the spread in gain between different tubes
suggests that a more robust option is  an increase in the linear
dynamic range of the front end electronics. This option has been
implemented in a new iteration of this ASIC, presently being
characterized.

\section{Conclusions}
We  successfully developed and characterized custom made ASICs to
process the signal from HPD and MaPMT photon detectors for the
BTeV RICH gas detector. Although the BTeV experiment was
terminated by an executive budgetary decision, we envisage a
variety of applications for these devices in high energy physics,
astrophysics and medical applications.

\section{Acknowledgements}
I would like to acknowledge the conference organizers for an
outstanding scientific program in an inspiring and charming
setting. Many thanks are due to my colleagues of the BTeV RICH
group: S. Blusk, C. Boulahouache, J. Butt, O. Dorjkhaidav, N.
Menaa, R. Mountain, H. Muramatsu, R. Nandakumar, L. Redjimi, K.
Randrianarivony, T. Skwarnicki, S. Stone, R. Sia, J. Wang, and H.
Zhang. I would like to thank S. Mikkelsen and B. Sundal, and the
other IDEAS engineers, whose electronics skills made this
development possible. I would also like to thank my BTeV
collaborators, for challenging discussions in the very productive
years of this research and development. Finally, my warmest thanks
are due to our spokespersons, J. Butler and S. Stone, whose
dedication to the success of BTeV was a true inspiration. This
work was supported by the US National Science Foundation.
\label{intro}

 \appendix

\end{document}